\documentclass [12pt,a4paper,leqno]{article}

\def\R{\mathrm{R}}

\title{Stationary Cylindrical Symmetric Solution approaching
 Einstein's cosmological solution}
\author{M.D.IFTIME\thanks{Queen Mary College, London E1 4NS, England}}
\date{26th November 2001}

\begin{document}

\maketitle

\begin{abstract}
Here we describe a stationary cylindrically symmetric 
solution of Einstein's equation with matter consisting of a positive 
cosmological and rotating dust term. The solution approaches
Einstein static universe solution.
\end{abstract}

\section{Introduction}
Similar problems for Einstein's equation without cosmological constant
and with negative cosmological constant are already solved in 
literature.
In ([1]) a rotating dust cylinder cut out of a Godel universe is
matched at exterior to a vacuum stationary cylindrically symmetric solution 
with negative cosmological constant.
In [3] Van Stockum found a rigidly rotating infinitely dust cylinder
without cosmological constant which has various exterior metrics.

In this paper we study cylindrically symmetric solutions of the 
Einstein's field equation with dust and positive cosmological constant which 
approaches Einstein static universe.

The spatially closed, static Einstein universe in usual form,

\begin{eqnarray} \label{eq:E}
\mathrm{d}s_E^2=\mathrm{d}\eta ^2+\sin ^2\eta(\mathrm{d}\theta ^2+\sin ^2\theta
\mathrm{d}\varphi ^2)-c^2\mathrm{d}\psi ^2\\
\varphi\in[0,2\pi ],\quad \eta\in [0,\pi ],\quad \theta\in [0,\pi ],
\quad \psi \in \R \nonumber
\end{eqnarray}

is the simplest cosmological dust model with constant curvature
 $K=\mathrm{const}$ and positive cosmological constant 
$\Lambda =\textrm{const}$, $\Lambda>0$. The field is produced by a 
energy-momentum tensor $T_{ab}$ of perfect-fluid:

\begin{equation}\label{eq:momentum}
\kappa T_{ab}=-\Lambda g_{ab}+\mu u_a u_b,\quad \mu>0,
\Lambda =\textrm{const.}>0.
\end{equation}

where $\Lambda={1\over K^2}$ and $\mu={2\over K^2}=2\Lambda=const.$.

For the exterior, we will use quite extensively the Einstein metric in 
cylindrical coordinates:

\begin{equation}\label{Einstein}
\mathrm{d}s^2=e^{2V_{0}(r)}(\mathrm{d}r ^2+\mathrm{d}z^2)+
W_{0}^{2}(r)\mathrm{d}\varphi ^2 - \mathrm{d}t^2
\end{equation}

where $W_{0}(r)$ and $V_{0}(r)$ have the form \cite{mt:SpaceTime}:

\begin{equation}
 V_{0}(r)={1\over 2}\mathrm{ln} \Bigl( \gamma-\lambda^2 \Biggl( \frac{1-e^{2\lambda (r-\nu)}}
{1+e^{2\lambda (r-\nu)}} \Biggr)^2 \Bigr) - \mathrm{ln}\sqrt{\Lambda}
\end{equation}

\begin{equation}\label{Wzero}
W_{0}(r)=\frac{1-e^{2\lambda (r-\nu)}}{(1-e^{2\lambda (r-\nu)})^{\gamma \over \lambda^3}}\,
e^{2\lambda (r-\nu)({1\over 2}-{\gamma \over 2\lambda^3})}
\end{equation}

where $\gamma$, $\alpha$, $\lambda\neq0$ and $\nu$ are constants of integration
and $\mu = 2\Lambda = \mathrm{const}$, the dust density respectively.

The space-time of Special Relativity is described mathematicaly by the
Minkowski space $(M,\, \eta)$. 
The flat metric $\eta$ in spherical coordinates \footnote{The coordiantes are 
singular in $r = 0$ and $\sin \theta = 0$.} has the form:
\begin{eqnarray} \label{eq:M}
\mathrm{d}s^2=\mathrm{d}r^2+r^2(\mathrm{d}\theta ^2 +\sin^2\theta
\mathrm{d}\varphi ^2 ) - c^2\mathrm{d}t^2 \\
r\in [0, \infty), \quad \theta \in [0, \pi], \quad \varphi \in [0, 2\pi], \quad
t \in \R, \nonumber
\end{eqnarray}
Minkowski spacetime is conformal to a finite region of the 
Einstein static universe
(\ref{eq:E}).

We have:
\begin{equation} \label{eq:tr}
 \mathrm{d}s^2 = \Theta^{-2}\mathrm{d}s_E^2,
\end{equation}
where $\Theta = 2\cos\frac{\psi+\eta}{2}\cos\frac{\psi-\eta}{2}$ is a smooth
strictly positive function, under the conformal transformation:

$$t+r = \mathrm{tg}\frac{\psi+\eta}{2}, \enspace t-r = \mathrm{tg}\frac{\psi-
\eta}{2}, \enspace -\frac{\pi}{2}\leq \psi-\eta \leq \psi+\eta \leq
\frac{\pi}{2}$$
(the boundaries $\psi\pm\eta = \pm\frac{\pi}{2}$ are the null surfaces
$\mathcal{I}^+$ and $\mathcal{I}^-$).

The de Sitter space-times are also conformal to a (finite) part of
$\mathrm{d}s_E^2$ and generally, all the closed Robertson-Walker metrics
(Minkowski space, the Sitter space are included as special cases) are
conformally equivalent to the Einstein static universe.
The Robertson-Walker metric,
\begin{equation}
\mathrm{d}s_{R-W}^2=K^2(ct)[\mathrm{d}\eta ^2+\sin ^2\eta (\mathrm{d}\theta^2 +
\sin ^2\theta \mathrm{d}\varphi ^2)]-c^2\mathrm{d}T^2
\end{equation}
under the transformation
\begin{equation} \label{eq:tbeg}
 \mathrm{d}T = \frac{1}{K(ct)}\mathrm{d}\psi
\end{equation}
we obtain the Einstein static universe metric (\ref{eq:E}) .


\section{The metric}

Stationary gravitational fields are characterized by the existence of a
 timelike Killing vector field $\xi$. Therefore in a stationary space-time
 $(M,\,g)$ we can construct a global causal structure. In other words we 
can introduce a coordinate system
$(x^a) = (x^{\alpha},\,t)$ with $\xi = \frac{\partial}{\partial t}$.
The metric $g_{ab}$ in these coordinates is independent of $t$ and has 
the general following form:
\begin{equation} \label{eq:stat}
 \mathrm{d}s^2 = h_{\alpha\beta}\mathrm{d}x^{\alpha}\mathrm{d}x^{\beta} +
F{(\mathrm{d}t + A_{\alpha}\mathrm{d}x^{\alpha})}^2, \qquad F \equiv \xi_a\xi^a
< 0.
\end{equation}
 
The unitary timelike vector field
$h^0 \equiv {(-F)}^{-{1\over 2}}\xi$ is globally defined on $M$
indicating the time-orientation 
in every point $p \in M$. Also it gives a global time coordinate
$t$ on $M$.(see [5])

Stationarity (i.e. time translation symmetry) implies that there exists a
1-dimensional group $G_1$ of isometries $\phi_t$ whose orbits are timelike
curves parametrized by $t$.

Using the 3-projection formalisme (developed by Geroch (1971)) of a
 4-dimensional spacetime manifold $(M,\,g)$
onto the 3-dimensional differentiable factor manifold 
 $\mathcal{S}_3=M/G_{1}$,
 the Einstein's field equations:
\begin{equation}
 R_{ab} - {1\over 2} R g_{ab} = \kappa T_{ab},
\end{equation}
for stationary fields take the following simplified form:
\begin{equation} \label{eq:einstein}
\left \{
\begin{array}{l}
\displaystyle R^{(3)}_{ab}={1\over 2}F^{-2}(\frac{\partial F}{\partial
x^a}\frac{\partial F}{\partial x^b}+\omega_{a}\omega_{b})
+\kappa(h^c_ah^d_b-F^{-2}\tilde{h}_{ab}\xi^a\xi^b)(T_{cd}-{1\over 2}Tg_{cd});

\vspace{6pt} \\ \vspace{6pt}

\displaystyle F^{\parallel a}_{,a}= F^{-1}\tilde{h}_{ab}(\frac{\partial
F}{\partial x^a}\frac{\partial F}{\partial x^b}-\omega_{a}
\omega_{b})-2\kappa F^{-1}\xi^a\xi^b(T_{ab}-{1\over 2}Tg_{ab});\\
\displaystyle \omega^{\parallel a}_a=2F^{-1}\tilde{h}_{ab}\frac{\partial
F}{\partial x^a}\omega_b

\vspace{6pt} \\ \vspace{6pt}

\displaystyle F \epsilon^{abc} \omega_{c,b} = 2 \kappa h_b^a T_c^b \xi^c
\end{array}
\right .
\end{equation}

Here, ``$\parallel$'' denotes the covariant derivative associated with the
conformal metric tensor $\tilde{h}_{ab} = -F h_{ab}$ on $\mathcal{S}_3$
($h_{ab} = g_{ab} + h^0_a h^0_b$ is the projection tensor) and $\omega^a =
{1\over 2}\epsilon^{abcd} \xi_{b} \xi_{c;d} \neq 0$ \footnote{Here I use the
convention: round brackets denote symmetrization and square brackets
antisymmetrization and $\Omega$ is the angular velocity.} is the rotational vector
($\omega^a\xi_a=0,\enspace \pounds_\xi\omega=0$).

We shall consider that the metric $g_{ab}$ has a cyllindrical symmetry, i.e,
it admits as well an Abelian group of isometries $G_2$ generated by 
two spacelike Killing vector fields $\eta$ and
$\zeta$, $\pounds_{\eta}g_{ab} = \pounds_{\zeta}g_{ab} = 0, \enspace
\eta_a\eta^a > 0, \enspace \zeta_a\zeta^a > 0$ and the integral curves of
$\eta$ are closed (spatial) curves.

There is a theorem (Kundt) which states that an axisymmetric metric can
be written in a (2+2)-split if and only if the conditions:
\begin{equation}
(\eta^{[a}\xi^b\xi^{c;d]})_{;e}=0=(\xi^{[a}\eta^b\eta^{c;d]})_{;e}
\end{equation}
are satisfied.

The existence of the orthogonal 2-surfaces is assured for the dust 
solutions, provided that the 4-velocity of dust satisfies the condition:
\begin{eqnarray}
u_{[a}\xi_b\eta_{c]}=0,\quad u^a=(-H)^{-{1\over 2}}(\xi^a+\Omega\eta^a)=
(-H)^{-{1\over 2}}l^i\xi^a_i,\quad\textrm{where}\\
l^i\equiv(1,\Omega),\quad H=\gamma_{ij}l^il^j,
\quad \gamma_{ij}\equiv \xi^a_i\xi_{aj},\quad i,j=1,2,\quad \xi_1=\xi;\ \xi_2=\eta \nonumber
\end{eqnarray}
)
in other words if the trajectories of the dust lie on the 
transitivity surfaces of the group
generated by the Killing vectors $\xi$, $\eta$.
In what follows, we will assume that this is true.
Using an adapted coordinate system, the metric (\ref{eq:stat})
 can be written in standard form:
\begin{equation} \label{eq:dust}
\mathrm{d}s^2=e^{-2U}[e^{2V}(\mathrm{d}r
^2+\mathrm{d}z^2)+W^2\mathrm{d}\varphi ^2]-
e^{2U}(\mathrm{d}t+A\mathrm{d}\varphi )^2
\end{equation}
where the functions \footnote{The function $W$ is defined invariantly as
$W^2 \equiv -2\xi_{[a}\eta_{b]}\xi^a\eta^b$.} $U$, $V$, $W$ and $A$
depend only on the coordinates $(r,z)$; these coordinates are also conformal
flat coordinates on the 2-surface $S_2$ orthogonal to 2-surface $T_2$ of the
commuting Killing vectors $\xi = \partial_t$ and $\eta = \partial_\varphi$.

If we identify the 4-velocity of the dust $u^a$ with timelike Killing vector
$\xi^a = \partial_t = (0, 0, 0, 1)$ then (\ref{eq:dust}) represents a co-moving
system $(x^1=r,\ x^2=z,\ x^3=\varphi,\ x^0=t)$ with dust,
$u_a = \xi_a = (0,\, 0,\, -e^{2U}A,\, -e^{2U})$ and
\begin{equation} \label{eq:metric}
\left \{
\begin{array}{l}
g_{11}=g_{22}=e^{-2U+2V}=h_{11}=h_{22},\\
g_{33}=e^{-2U}W^2-e^{2V}A^{2}=h_{33},\quad g_{00}=\xi_0=-e^{2U}=F,\\
g_{03}=\xi_3=-e^{2U}A,\quad g_{13}=g_{23}=g_{10}=g_{20}=0
\end{array}
\right .
\end{equation}

We can use the complex coordinates $(q, \bar{q})$ on the 2-surface $S_2$:
\begin{equation}
 q={1\over \sqrt{2}}(r+iz)
\end{equation}
 and the stationary axisymmetric metric (\ref{eq:dust}) takes the Lewis-
Papapetrou form:

\begin{equation} \label{eq:dust2}
\mathrm{d}s^2=e^{-2U}(e^{2V}\mathrm{d}q\mathrm{d}\bar{q}+
W^2\mathrm{d}\varphi ^2)-
e^{2U}(\mathrm{d}t+A\mathrm{d}\varphi )^2
\end{equation}

The surface element on $T_2$ is $f_{ab}=2\xi_{[a}\eta_{b]}$, $f_{ab}f^{ab}<0$
and the surface element on $S_2$ is
$\tilde{f}_{ab}$, the dual tensor of $f_{ab}$,
$\tilde{f}_{ab}={1\over 2}\epsilon_{abcd}f^{cd}$

Thus the Einstein's dust equations with constant $\Lambda > 0$
(\ref{eq:einstein}) for the metric (\ref{eq:metric}) will take the following
form:
\begin{equation}\label{eq:einstein2}
\left \{
\begin{array}{l}
\displaystyle
\frac{\partial^{2}W}{\partial q\partial \bar{q}}=-\Lambda We^{2V-2U}
\vspace{6pt} \\ \vspace{6pt}
\displaystyle
\frac{\partial^{2}U}{\partial q\partial \bar{q}}+\frac{1}{2W}(\frac{\partial
U}{\partial q} \frac{\partial W}{\partial \bar{q}}
+\frac{\partial U}{\partial \bar{q}} \frac{\partial W}{\partial
q})+\frac{1}{2W^2}e^{4U} \frac{\partial A}{\partial q}
\frac{\partial A}{\partial \bar{q}} = (\mu-2\Lambda)\frac{e^{2V-2U}}{4}
\vspace{6pt} \\ \vspace{6pt}
\displaystyle
\frac{\partial^{2}A}{\partial q\partial \bar{q}}-\frac{1}{2W}(\frac{\partial
A}{\partial q} \frac{\partial W}{\partial \bar{q}}
+\frac{\partial A}{\partial \bar{q}} \frac{\partial W}{\partial q}) +
2(\frac{\partial A}{\partial q} \frac{\partial U}{\partial \bar{q}}
+\frac{\partial A}{\partial \bar{q}} \frac{\partial U}{\partial q})=0
\vspace{6pt} \\ \vspace{6pt}
\displaystyle
\frac{\partial^{2}W}{\partial q\partial \bar{q}}-2\frac{\partial W}{\partial q}
\frac{\partial V}{\partial q}+
2W(\frac{\partial U}{\partial q})^2-\frac{1}{2W}e^{4U}(\frac{\partial
A}{\partial q})^2=0
\vspace{6pt} \\ \vspace{6pt}
\displaystyle
\frac{\partial^{2}V}{\partial q\partial \bar{q}}+\frac{\partial U}{\partial q}
\frac{\partial U}{\partial \bar{q}}
+\frac{1}{(2W)^2}e^{4U}\frac{\partial A}{\partial q} \frac{\partial A}{\partial
\bar{q}}=
 -\Lambda \frac{e^{2V-2U}}{2}
\end{array}
\right .
\end{equation}
Here $\displaystyle\Delta=\frac{\partial^2}{\partial r^2} +
\frac{\partial^2}{\partial z^2}=2\frac{\partial^2}{\partial q \partial\bar{q}}$
is the Laplace operator and the energy-momentum tensor $T_{ab}$ has the form
(\ref{eq:momentum}) with $\Lambda=const.>0$ and $\mu(r)>0$.

The conservation law $T_{;b}^{ab}=0$ implies $U_{,a}=0$. We obtain then
$ U=const.$  a consequence
of the field equations which will use it in place of one of the Einstein's 
equations.

Moreover assuming that $U=0$ in the expressions of metric functions
(\ref{eq:metric}) and we obtain that the matter
 current paths are geodesics
($\dot{u}_a=u_{a;b}u^b=0$), without expansion ($\theta=u_{;a}^a=0$),
in a non-rigidly rotation 
($\omega=\sqrt{{1\over 2}\omega_{ab}\omega^{ab}} \neq 0$)
and with $\sigma \neq 0$.

The field equations (\ref{eq:einstein2}) will take then the following 
simplified form:
\begin{equation} \label{eq:einstein3}
\left \{
\begin{array}{l}
\displaystyle
\frac{\partial^{2}W}{\partial q\partial \bar{q}}=-\Lambda We^{2V}

\vspace{6pt} \\ \vspace{6pt}

\displaystyle
\frac{1}{2W^2} \frac{\partial A}{\partial q} \frac{\partial A}{\partial \bar{q}}
 = (\mu-2\Lambda)\frac{e^{2V}}{4}

\vspace{6pt} \\ \vspace{6pt}

\displaystyle
\frac{\partial^{2}A}{\partial q\partial \bar{q}}-\frac{1}{2W}(\frac{\partial
A}{\partial q} \frac{\partial W}{\partial \bar{q}}
+\frac{\partial A}{\partial \bar{q}} \frac{\partial W}{\partial q}) = 0

\vspace{6pt} \\ \vspace{6pt}

\displaystyle
\frac{\partial^{2}W}{\partial q\partial \bar{q}}-2\frac{\partial W}{\partial q}
\frac{\partial V}{\partial q}-
\frac{1}{2W}(\frac{\partial A}{\partial q})^2=0

\vspace{6pt} \\ \vspace{6pt}

\displaystyle
\frac{\partial^{2}V}{\partial q\partial \bar{q}}
+\frac{1}{(2W)^2}\frac{\partial A}{\partial q} \frac{\partial A}{\partial
\bar{q}}= -\Lambda \frac{e^{2V}}{2}
\end{array}
\right .
\end{equation}

Taking into account the third symmetry i.e., the presence of the 
spacelike Killing vector field $\zeta = \partial_z$
we reduced to the problem of solving the
Einstein's system of ordinary differential equations (\ref{eq:einstein3})
for the metric:
$$  
\mathrm{d}s^2=e^{2V(r)}(\mathrm{d}r^2 + \mathrm{d}z^2) +
W^2(r)\mathrm{d}\varphi - (\mathrm{d}t+A(r)\mathrm{d}\varphi)^2
$$
in the unknown metric functions 
$V(r)$, $W(r)$, $A(r)$ and $\mu(r)$ 
and to match the constant of integration such that the exterior 
field is conformal with Einstein static universe 
\footnote{For the case when $\Lambda = 0$  has an exterior static
($\omega^a\!=\!{1\over 2}\epsilon^{abcd}\xi_b\xi_{c;d}\!=0$) even the dust was
in rotation with $\Omega = \textrm{const.}$ (Van Stockum class solutions).}

If we denote ${\partial\over \partial r}='$ the field equations
(\ref{eq:einstein3}) will take the simplified form:
\begin{equation} \label{eq:einstein4}
\left \{
\begin{array}{l}
\displaystyle W''=-2\Lambda We^{2V} \\
\displaystyle 2{A'}^2=(\mu-2\Lambda)W^2e^{2V} \\
\displaystyle \frac{A''}{A'} = 2\frac{W'}{W} \\
\displaystyle W''-4W'V'-{1\over W}A'^2=0 \\
\displaystyle V''+{1\over 2W^2}{A'}^2=-\Lambda e^{2V}
\end{array}
\right .
\end{equation}

The system of equations (\ref{eq:einstein4}) can be further reducible
 to the following form:

\begin{equation} \label{eq:einstein5}
\left \{
\begin{array}{l}
\displaystyle W''=-2\Lambda We^{2V} \\
\displaystyle A'=aW^2 \\
\displaystyle (\mu-2\Lambda)e^{2V}= 2a^{2}W^2 \\
\displaystyle W''-4W'V'-a^{2}W^3 =0 \\
\displaystyle W''-a^{2}W^{3}-2 V''W=0
\end{array}
\right .
\end{equation}
where $a\neq 0$,  $b\neq 0$ are positive constants
and $\Lambda$ is the positive cosmological constant, which is very small
(less than $10^{-57} cm^{-2}$).

After further simplifications the system of equations (\ref{eq:einstein5})
will take the form:

\begin{equation} \label{eq:einstein6}
\left \{
\begin{array}{l}
\displaystyle \frac{W''}{W}=-2\Lambda e^{2V} \\
\displaystyle A'=aW^2 \\
\displaystyle V'=bW^2 \\
\displaystyle (\mu-2\Lambda)e^{2V}= 2a^{2}W^2 \\
\displaystyle W''-4bW'W^{2}-a^{2}W^3=0
\end{array}
\right .
\end{equation}

where $a$ and $b$ are positive constants of integration.

The system (\ref{eq:einstein6}) does not have an explicit analytical 
solution for $W(r)$, $V(r)$, $A(r)$ and $\mu(r)$ as functions of radius $r$.
Therefore we shall look to derive a good approximation of the solution.

We remark from the form of the system, that we 
are looking for a  one-parameter\footnote{It is actually a two-parameter
family of solutions $g_{ij}(a,b)$} family $g_{ij}(a)$ of solutions, where $a$ 
measures the size of perturbation, in the sense that $g_{ij}(a)$ are
continuous differentiable on $a$ and for $a=0$ we obtain
Einstein universe solution.

In what follows we shall show that the the solution of (\ref{eq:einstein6})  
is approaching Einstein universe solution $ g_{ij}(r,0)$, as radius $r$ goes 
to zero.

\begin{equation}
 g_{ij}(r,a) = g_{ij}(r,0) + a g_{ij,a}(r,0) + a^2 g_{ij,aa}(r,0) +...
\end{equation}

We shall perturb the solution as power series in $a$ 
about Einstein universe solution and give a good approximation to 
$g_{ij}(a)$ for sufficiently small $a$. 

To do so we  differentiate the system (\ref{eq:einstein6}) with 
respect to $a$, then take $a$ to be zero and obtain the following equations:

\begin{equation} \label{eq:einstein7}
\left \{
\begin{array}{l}
\displaystyle \dot{W''}(r,0)- P(r)\dot{W'}(r,0)- Q(r)\dot{W}(r,0)=0 \\
\displaystyle \dot A'(r,0)=W_{0}^2 \\
\displaystyle \dot{\mu}(r,0)e^{2V_{0}}=0\\
\displaystyle \dot{V'}(r,0)=2b W_{0}\dot W(r,0)
\end{array}
\right .
\end{equation}

for the functions $\dot {W}(r,0)$, $\dot {V}(r,0)$, $\dot {\mu}(r,0)$ 
$\dot{A}(r,0)$.

We denoted by ${\partial \over \partial a}=\dot{}$ \hspace{.05in},
$P(r)=4bW_{0}^2$,\hspace{.1in}$Q(r)=8bW_{0}W'_{0}$
and $W_{0}(r)$, $V_{0}(r)$ are the metric functions of
Einstein universe solution (\ref{Einstein}). 

By choosing appropriate constants of integration
$\nu=0$, $\lambda=1$, $\gamma=1$ in (\ref{Wzero})
we get $W_{0}(r)=1$. \cite{mt:SpaceTime}

Then the  system (\ref{eq:einstein7}) can be completely integrated and take 
the following form:

\begin{equation} \label{eq:einstein8}
\left \{
\begin{array}{l}
\displaystyle \dot{W}(r,0)=c_1 +c_{2}e^{4br}\\
\displaystyle \dot {A}(r,0)=r(1+2ac_{1})+\frac{ac_2}{2b}e^{4br}\\
\displaystyle \dot {\mu}(r,a)=0\\
\displaystyle \dot V(r,0)=2bc_{1}r +\frac{c_2}{2}e^{4br}
\end{array}
\right .
\end{equation}

Then $g_{ij}(r,0) + a g_{ij,a}(r,0)$ will give a good approximation to
solution of the Einstein field equations (\ref{eq:einstein6}) 
$g_{ij}(r,a)$ for small $a$ when $r$ approaches 
the axis of rotation $\eta=0$.

\begin{equation} \label{eq:lasteinstein}
\left \{
\begin{array}{l}
\displaystyle W(r,a)=1 + ac_1 + ac_{2}e^{4br}\\
\displaystyle A(r,a)= ar(1+2ac_{1})+\frac{a^{2}c_2}{2b}e^{4br}\\
\displaystyle \mu(r,a)=2\Lambda \\
\displaystyle V(r,a)= V_{0}(r)+ 2abc_{1}r +\frac{ac_2}{2}e^{4br}
\end{array}
\right .
\end{equation}

\section{Conclusion}
In this paper we have been studying a spacetime satisfying Einstein field
equations with positive cosmological constant, describing a dust cylinder in 
non-rigid rotation, which approaches Einstein's cosmological static solution
on the axis of rotation.
The metric is given in approximation around the axis of rotation and it 
depends on three parameters $a$, $b$ and $\Lambda$.

\section{Acknowledgements}

I would like to thank Professor Malcom MacCallum for encouragement and
precious advice. I would also like to thank Professor W B Bonnor 
for reading the paper and making useful comments, Dr Thomas Wolf and Mr 
Ady Penisoara for help in using computer programs.
Last but not least I would like to thank my fiance, Joe for patience and 
support.

\end{document}